%% file: morello_acat2019.tex
\documentclass[a4paper]{jpconf}
\usepackage{graphicx}
\usepackage{hyperref}

\usepackage{lineno}

\input{lhcb-symbols-def_mjm}

\input{my-symbols-def}

\begin{document}

\title{Real-time reconstruction of long-lived particles at LHCb using FPGAs}

\author{Riccardo~Cenci$^{1,2}$, Andrea~Di~Luca$^{1,3}$, Federico~Lazzari$^{1,4}$, Michael~J.~Morello$^{1,2}$, Giovanni~Punzi$^{1,3}$\\ on behalf the LHCb Collaboration}

\address{
$^1$ INFN sezione di Pisa, Largo B. Pontecorvo 3 - 56127 Pisa, Italy\\
$^2$ Scuola Normale Superiore, Piazza dei Cavalieri 7- 56126 Pisa, Italy\\
$^3$ Universit\`a di Pisa, Lungarno Pacinotti 43 - 56126 Pisa, Italy\\
$^4$ Universit\`a degli Studi di Siena, via Banchi di Sotto 55 - 53100 Siena, Italy}

\ead{michael.morello@sns.it, michael.morello@pi.infn.it}

\begin{abstract}
Finding tracks downstream of the magnet at the earliest LHCb trigger level is not part of the baseline plan of the upgrade trigger, 
on account of the significant CPU time required to execute the search. Many long-lived particles, such as \KS and strange baryons, decay after the vertex track detector, so that their reconstruction efficiency is limited. We present a study of the performance of a future innovative real-time tracking system based on FPGAs, developed within a R\&D effort in the context of the LHCb Upgrade Ib (LHC Run~4), dedicated to the reconstruction of the particles downstream of the magnet in the forward tracking detector (Scintillating Fibre Tracker), that is capable of processing events at the full LHC collision rate of 30 MHz.
\end{abstract}

\section{Introduction}
 
The \lhcb detector, collected data at a luminosity of $4 \times 10^{32} $cm$^{-2}$s$^{-1}$ until the end of LHC Run~2. 
During the Long Shutdown 2 (LS2) it will be replaced by an upgraded experiment, referred as the Phase-I Upgrade (LHC Run~3, 2021-2024 and LHC Run~4, 2027-2029). The Phase-I Upgrade, operating at a 
luminosity of $L= 2 \times 10^{33} $cm$^{-2}$s$^{-1}$, will greatly improve the sensitivity of many flavour studies. However, the precision on a host 
of important, theoretically clean, measurements will still be limited by statistics, and other observables associated with highly suppressed 
processes will be poorly known. There is therefore a strong motivation for a consolidation of the the Phase-I Upgrade in view of the LHC Run~4,
and for building a Phase-II Upgrade, which will fully realize the flavour potential of the High-Luminosity LHC during the LHC Run 5 ($\geq 2031$) at a luminosity $L > 10^{34} $cm$^{−2} $s$^{-1} $~\cite{LHCb-PII-EoI,TheLHCbCollaboration:2320509}. 

Although the trigger strategy of both the Phase-I and Phase-II Upgrades is software based, studies are underway to learn what benefits could accrue 
by adding dedicated processors to help solve specific low-level tasks. One relevant example is to find tracks downstream of the magnet 
at the earliest trigger level~\cite{LHCb-PII-EoI,TheLHCbCollaboration:2320509}. This capability is not part of the baseline trigger scheme on account of the significant CPU time 
required to execute the search. Not having access to this information greatly limits efficiency for decay modes with downstream tracks that cannot easily 
be triggered through another signature, for example channels containing a \KS  and less than two prompt charged hadrons, as $B\to \KS\KS$, $B\to \KS\KS\KS$, $B\to \eta’\KS$, $B\to \phi \KS$, $B\to \omega \KS$, 
$D^0\to \KS\KS$, $\Dsp \to \KS \pi^+$, $\Dp \to \KS K^+$, $ \KS\to \mu \mu$, etc. The same is 
true for decays involving $\Lambda$ baryons (i.e. $\Lambda^0_b \to 3\Lambda$) and long-lived exotic particles (hidden sector WIMP Dark Matter and Majorana neutrinos). 

An  R\&D work is currently ongoing~\cite{simone_ttfu_talk}, within the LHCb Collaboration, for the  realization of   
an innovative tracking device, the so-called Downstream Tracker, capable of 
reconstructing in real time long-lived particles in the context of the envisioned future upgrades (beyond LHC Run~3) of the \lhcb experiment, with the aim of  recovering 
the reconstruction efficiency of the downstream tracks.  Such a specialized processor is supposed to obtain a copy of the data from the readout system, reconstruct downstream tracks, 
and insert them back in the readout chain before the event is assembled, in order to be sent to the high level trigger
in parallel with the raw detector information. This approach, where tracks can be seen as the output of an additional  ``embedded track detector''
is based on the \textit{artificial retina} algorithm~\cite{Ristori:2000vg,lltt}, which is a highly-parallel pattern-matching algorithm, 
whose architectural choices, inspired to the early stages of image processing in mammals, make 
it particularly suitable for implementing a track-finding system in present-day FPGAs. 
First small prototypes of the track-processing unit, able to reconstruct two-dimensional straight-line tracks in a 6-layers realistic tracking detector,  
based on the artificial retina algorithm have been designed, simulated, and built\cite{Stracka:2017,lazzari_master_thesis,cenci_proc_2017,lazzari_proc_2018} using commercial boards, equipped with modern 
high-end  FPGAs.  Throughputs for processing realistic LHCb-Upgrade\footnote{Running conditions in Run~4 will be the same as in Run~3, the so-called LHCb-Upgrade.} events above 30~MHz and latencies lesser than 1~$\mu \textrm{s}$  
have been achieved running at the nominal clock speed, demonstrating the feasibility of fast track-finding with a FPGA-based system.
This opens the way for a full realistic application as the Downstream Tracker. 

\section{The tracking system of the LHCb Upgrade I}

The \lhcb detector~\cite{Alves:2008zz,Aaij:1978280} is a single-arm forward spectrometer covering the \mbox{pseudorapidity} range $2<\eta <5$, and 
it is specifically designed for the study of particles containing  \bquark- or \cquark-quarks.  
The \lhcb Upgrade detector~\cite{Bediaga:1443882}, has a similar layout of the previous experiment, and it  includes a high-precision tracking system 
consisting of a hybrid pixel sensors vertex detector (VELO) surrounding the \pp interaction region, a large-area silicon-strip detector (Upstream Tracker or UT) located 
upstream of a dipole magnet with a bending power of about $4{\mathrm{\,Tm}}$, and three stations of scintillating fibres detectors 
(Scintillating Fibre Tracker or SciFi)\cite{Collaboration:164740} placed downstream of the magnet. 
The SciFi, which is the subdetector used in the studies reported here, has three stations (T-stations) which are composed of four detection layers with a $x-u-v-x$ geometry, with vertical fibres in first and last layers, and tilted fibres by a stereo angle of $-5^{\circ}$ and of $+5^{\circ}$ in central layers.\footnote{LHCb uses a right-handed coordinate system  with the $z$ coordinate along the beam axis, and the $y$ coordinate along the vertical coinciding with the direction of the magnetic field.} 
%
The nominal spatial hit resolution is 72~$\mu \textrm{m}$, while the hit efficiency is 97.5\%.

The reconstructed tracks are divided into different types depending on the subdetectors in which they are reconstructed. 
The most valuable tracks for physics analysis are the so-called long tracks which are reconstructed in the VELO and the T-stations. They have excellent spatial resolution close to the primary interaction and a precise momentum information due to the combined information of the track slope before and after the magnet. Tracks consisting of measurements in the T-stations alone are known as T-tracks. They are not used in physics analyses, but are used as inputs to reconstruct the so-called downstream tracks. These are tracks which have measurements in the UT and the T-stations. They are important for the reconstruction of the daughters of long-lived particles such as 
\KS mesons or \Lz baryons which decay outside the VELO. Tracks consisting of measurements in the VELO and in the UT  are called upstream tracks, while
the so-called VELO tracks consist of measurements in the VELO only.  


\subsection{The seeding algorithm}
The reconstruction of downstream tracks is a well known challenge because of the much-higher combinatorial in the T-stations
with respect to the vertex tracker.
The algorithm performing a standalone track search in the T-stations is called \textit{seeding}, and it has to solve a complex and heavy pattern recognition task. 
At the current state-of-the-art it requires a significant amount of CPU-time to be executed, about a few hundred microseconds per event\footnote{Measured using a setup different from the official throughput test setup.}\cite{LHCb-PUB-2017-005}, while the total budget for the LHCb-Upgrade tracking sequence, in Run 3, is expected to be 33 $\mu$s per event, assuming 1000 Event Filter Farm nodes~\cite{LHCb-PUB-2017-005,LHCB-TDR-016}. 
Further CPU-time is also needed to link back T-tracks and add UT hits, in order to find and reconstruct the downstream tracks. At the moment, finding standalone T-tracks at the earliest trigger level is therefore not part of the baseline trigger scheme. 

\section{The artificial retina architecture}

A detailed description of the artificial retina architecture, along with  an early evaluation of its performance, can be found elsewhere~\cite{Stracka:2017,lazzari_master_thesis,cenci_proc_2017,lazzari_proc_2018}.  Only a brief summary is reported here.
The tracking process is made with two main stages. In the first one (switching), all hits received from the different tracking detector layers are coordinate-transformed, and delivered to the appropriate location in a processing array, through a large custom-built 
switching network.  During this stage a significant duplication of informations can occur, requiring the use of a large bandwidth.   The second stage is performed by a large array of cells (processing engines), mapping the track parameter space.  Each cell evaluates an appropriate weighting function, similar to the analog excitation response of biological neurons, related to the distance  of each hit from a set of reference tracks. The array is endowed with
the capability of performing local cluster finding 
to determine the location of tracks and their parameter estimates in a completely parallel fashion  over the entire device, without wait states, thus ensuring  high throughput and low latency. The size of a such envisioned device it is affordable with already commercially available off-the-shelf FPGAs; a full 
realistic tracking system requires  about $10^5$ processing engines~\cite{lazzari_master_thesis,cenci_proc_2017}.


The Downstream Tracker has to be integrated inside the DAQ architecture of the LHCb Upgrade, and precisely into the Event Builder (EB)~\cite{LHCB-TDR-016} which receives data from the detector readout. The Event Builder consists in  a cluster of 500 rack-mounted PC nodes, where each of them mounts a readout board for receiving data from subdetectors and two network interfaces, one connected to the internal EB network for the event building stage, and one to send data to the Event Filter Farm for the high level trigger. In the current foreseen layout, the SciFi subdetector will send raw hits to about 150 EB nodes, one half will receive data from the 6 $x$-coordinate layers and one half from the 6 $u/v$-coordinate layers. Gathering data from a such large number of nodes requires an equally large number of devices to perform the switching function. 
For this reason each EB node of the SciFi subdetector has to be instrumented with a standard commercial PCIe card equipped with FPGAs (tracking board) receiving  a copy of data from the readout before the beginning of the event building stage. Each tracking board will host both functionalities of  the switching network and of the processing engines.  Each portion of data, distributed in different tracking boards, will be sent to all relevant engines, 
through a mesh network (patch panel) allowing the exchange of the hits between different tracking boards.
Each tracking board will return a subsample of reconstructed track candidates to the EB node to which it is connected, and the reconstructed tracks candidates will be added to the raw data collected by that specific node. At this point the event building process will proceed as usual and reconstructed track candidates will be treated by the system as raw track-hits from an additional ``embedded track detector''.

\section{Reconstruction of T-tracks in real time}
The reconstruction of standalone T-tracks is the first and the most expensive part of the reconstruction of downstream tracks. Therefore, 
in order to design and realize a fully operational Downstream Tracker\footnote{The Downstream Tracker will use hits 
both from SciFi and UT subdetectors in order to reconstruct in real time downstream tracks.} this challenge has to be overcome at first. 
 These proceedings present, therefore, the first study of the performance of a real-time reconstruction of T-tracks, in the SciFi subdetector, achievable with the artificial retina architecture, using fully simulated events at the LHCb Upgrade (LHC Run~3 and 4) conditions. 

In order to develop the retina algorithm for the Downstream Tracker and measure its tracking performance three different simulated samples are used. They differ one from another only by the event topology. The first one  is a sample of generic inelastic events, the so-called Minimum Bias sample, while the other two are filtered samples containing an hard collision in each event, which has produced a $\Dstar^+ \to \Dz \pip \to  [\KS \pip\pim]\pip$ decay and a $B^0_s \to \phi\phi \to  [\Kp\Km][\Kp\Km]$ decay, respectively. The average number of reconstructible particles is about 150-200 particles per event, depending on the simulated sample. A non negligible tail of the distribution is present, reaching values up to 400.

The tracking problem is performed using a sequential approach with three main stages:
1)~find $x- z$ (or axial) track projection; 2)~removal of $x- z$ false positive tracks; and 3)~association of the stereo $u/v$-hits to the axial track candidates. 

The first stage is the most important since the pattern recognition task is mainly solved at this level.  Tracks are approximated as straight lines in this early reconstruction stage although the presence of a small component of  fringe field in the SciFi region,\footnote{This will result to be a very good approximation.} since  the artificial retina algorithm can be efficiently implemented on FPGAs only for a two-dimensional tracking problem.
The track axial projection is therefore 
parameterized using a the two-dimensional space ($x_0,x_{11}$), where $x_0$ and $x_{11}$ are the $x$-coordinates of the intersections of the track with two virtual planes located just before the first layer of the SciFi and just after the last one, respectively. This space is divided into about $10^5$ cells, corresponding precisely to 25800 cells per quadrant,\footnote{Quadrants can be considered independent for track reconstruction purpose, and all the tracking studies, done in these proceedings, are performed in a single quadrant.} where the region of interest is that around the diagonal, being populated by the majority of particles produced in the $pp$ interaction or from their subsequent decays. In order to avoid wasting resources only this region is therefore covered by the retina.  Each cell corresponds to a pattern track with parameters the center of the cell itself, which propagates into the physical space as a straight line,  where the geometric coordinates of the the intersections with the detector layers, called receptors, are pre-calculated and stored using the LHCb simulation.

\begin{figure}[!h]
\begin{center} 
\includegraphics[width=0.49\linewidth]{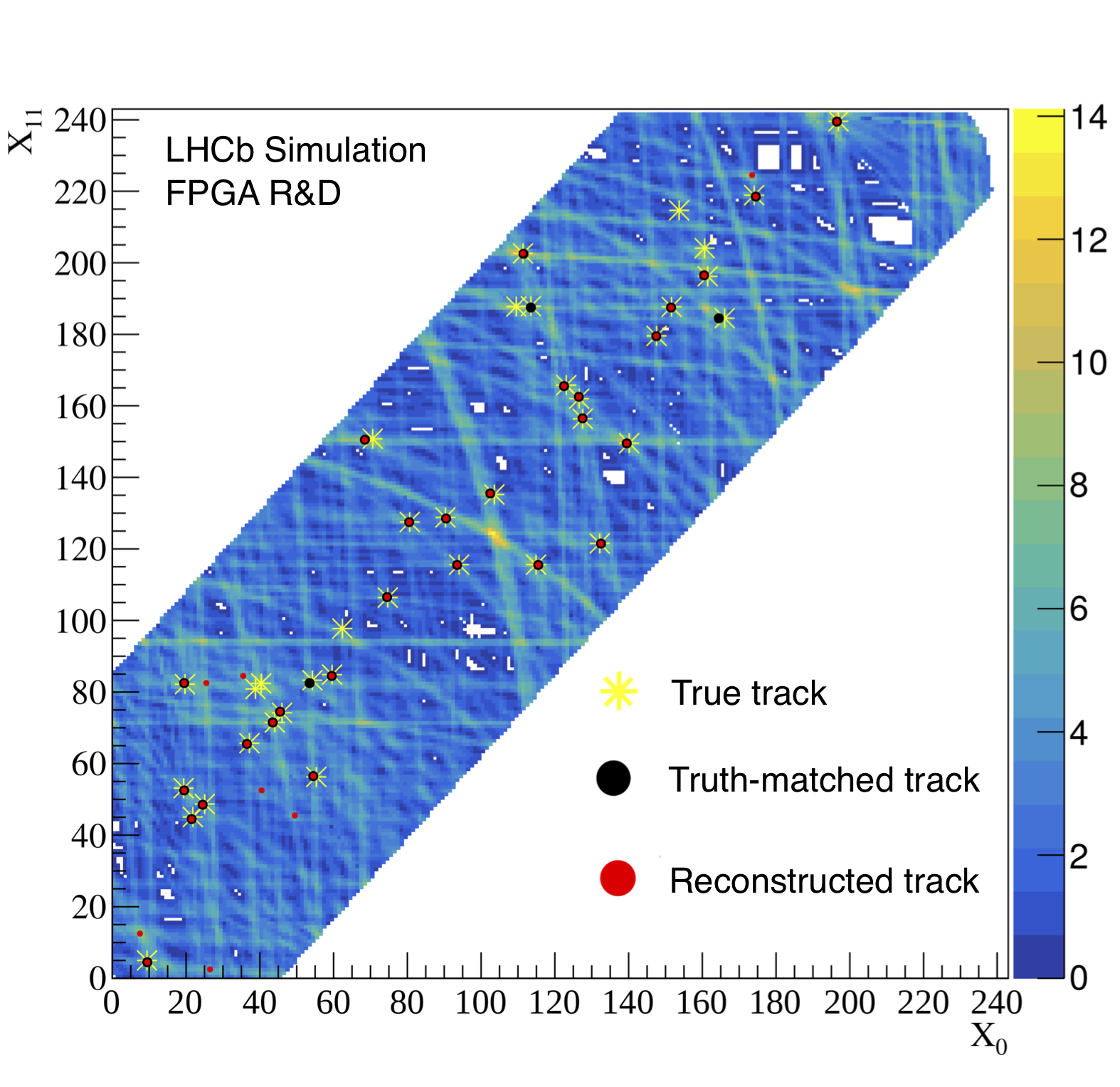}
\includegraphics[width=0.49\linewidth]{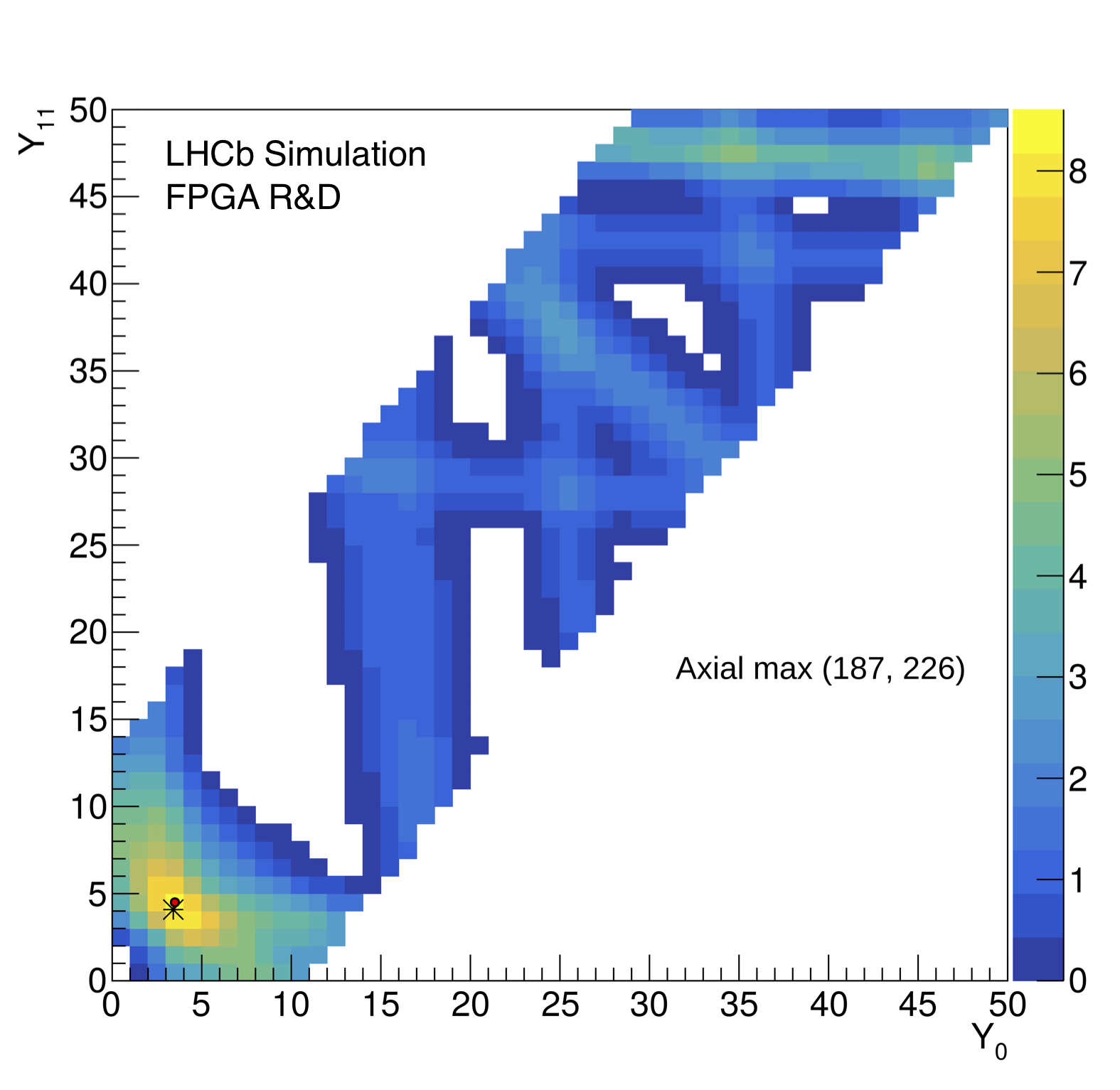}
\caption{Excitation level of the axial retina filled with \scifi subdetector hits from a single fully simulated event (left).
Excitation level of a stereo retina corresponding to a given axial local maximum (right).
True tracks (stars),  reconstructed track candidates (red dots), and truth-matched reconstructed track candidates (black dots). }
\label{fig:retinas}
\end{center} 
\end{figure}
SciFi detector hits, from realistic simulated \lhcb~Upgrade events,  are then sent to the axial retina. A gaussian weight is calculated and stored into the cell, only for the hits that are distant from the cell receptors less than a fixed distance. Hits are accumulated into the axial retina, and immediately after the end of each event, a search for local maxima is performed. Only maxima exceeding a certain excitation level threshold are promoted to be axial track candidates (find $x-z$ track projection). An efficiency well above the 95\% level is achieved for generic particles reconstructible in the SciFi subdetector\footnote{A Monte Carlo simulated particle is reconstructible in the SciFi subdetector if it has released at least one $x$-coordinate hit  and one $u/v$-coordinate hit in each of the three T-stations. The minimum number of hits for a track to be reconstructible in the SciFi subdetector  is therefore equal to six.}  with, however, a 90\% level of ``ghost rate'' of  fake track candidates.  The ghost rate is defined as the amount of reconstructed tracks not associated to a true Monte Carlo particle (truth-matching) with respect to the total amount of reconstructed track candidates. Such a level of fake tracks is not affordable, 
so an additional quality requirement is necessary to promote a local maximum as an axial track candidate. 
A threshold is therefore set on the $\chi^2_{\rm A}$ value, returned from a fit to 
the $x$-coordinate hits stored into the local maximum using a parabolic model.\footnote{The parabola accounts for the presence of the fringe field in the SciFi subdetector region to a good approximation.} 
The fit is performed over the combinations of  the two closest hits to the cell receptor for each layer.\footnote{The distribution of the number of combinations to be done has an average of 7, with a very small tail above 20. } This is made through a linearized fit strategy, suitable to be easily implemented on the DSP blocks of currently available FPGAs (removal of $x-z$ false positive tracks).
As an example, figure~\ref{fig:retinas} (on the left) shows the excitation level of the axial retina filled with the $x$-coordinates hits of the top right quadrant of the SciFi subdetector layers for a single event. The position of all the reconstructed track candidates is superimposed, along with 
the position of the true tracks, and with that of the truth-matched reconstructed tracks.
The resulting efficiencies in finding the axial track projection ($\varepsilon_{\rm A}$), shown in table~\ref{tab:efficiencies_3030_solostereo}, are close to the 90\% level for all the track categories if a minimal requirement on the track momentum is applied.\footnote{$p > 3, 5 \gev$ are minimal requirements that the majority of LHCb analyses uses for physics signal tracks.} The ghost rate drastically reduces to less than 20\%. Both tracking efficiencies and  ghost rate are comparable to those obtained with the offline reconstruction software program~\cite{Collaboration:164740,LHCb-PUB-2017-005}.
\begin{table}[!htbp]
\caption{Axial only  ($\varepsilon_{\rm A}$) and three-dimensional ($\varepsilon_{\rm AS}$) averaged reconstruction efficiencies for different simulated samples 
and different track categories. The ghost rate is also shown. The downstream strange tracks  are mainly pions from $\KS \to \pip\pim$ decay.}
\centering
    \resizebox{.8\textwidth}{!}{
      \begin{tabular}{l|cc|cc|cc}
      \br
                                 & \multicolumn{2}{c|}{\textrm{Minimum Bias}} & \multicolumn{2}{c|}{\textit{\Dz \to \KS\pip\pim}} & \multicolumn{2}{c}{\textit{$B^0_s  
\to \phi\phi$}}\\  
          Track type           & $\varepsilon_A$       & $\varepsilon_{AS}$ & $\varepsilon_A$       & $\varepsilon_{AS}$& $\varepsilon_A$       & $\varepsilon_{AS}$ \\
 \mr
T-track                                                                                    &   75.0     & 71.4     &    74.4      &   70.0   &   73.9     &      67.4     \\
T-track, $p > 3 \mathrm{GeV}/\mathrm{c}$                                                   &   87.0     & 83.0     &    85.9      &   80.8   &   85.1     &      77.2     \\
T-track, $p > 5 \mathrm{GeV}/\mathrm{c}$                                                   &   90.3     & 85.7     &    88.2      &   82.7   &   86.6     &      77.4     \\
Long                                                                                    &   81.7     & 78.8     &    84.1      &   79.5   &   84.2     &      77.2     \\
Long, $p > 3 \mathrm{GeV}/\mathrm{c}$                                                   &   87.3     & 84.2     &    87.1      &   82.3   &   87.3     &      79.8    \\
Long, $p > 5 \mathrm{GeV}/\mathrm{c}$                                                   &   90.6     & 86.9     &    88.1      &   83.1   &   88.1     &      79.9     \\
Downstream                                                                                    &   80.1     & 77.7     &    83.0      &   78.6   &   82.6     &      76.2     \\
Downstream, $p > 3 \mathrm{GeV}/\mathrm{c}$                                                   &   87.0     & 84.4     &    87.1      &   82.5   &   86.5     &      79.3     \\
Downstream, $p > 5 \mathrm{GeV}/\mathrm{c}$                                                   &   90.5     & 87.5     &    88.8      &   83.6   &   87.9     &      80.2     \\
Downstream strange                                                                            &   -        & -        &    84.7      &   82.8   &   -        &      -        \\
Downstream strange, $p > 3 \mathrm{GeV}/\mathrm{c}$                                           &   -        & -        &    89.4      &   86.7   &   -        &      -        \\
Downstream strange, $p > 5 \mathrm{GeV}/\mathrm{c}$                                           &   -        & -        &    93.0      &   87.2   &   -        &      -        \\
\mr                                                                                                                                                                                          
ghost rate                                                                               &   12.1     &  15.7     &   16.3      &   20.2   &   18.4     &      24.7     \\
\br                                                                                                                                  
\end{tabular}
}
\label{tab:efficiencies_3030_solostereo}
\end{table}%

Once the axial track candidates are found  the $u/v$-hits must be associated to each of them in order to find the relative $y-z$ track projection and reconstruct the three-dimensional T-tracks~\cite{diluca_master_thesis}.  The linearized fit to a parabola of the $x$-coordinate hits allows the determination of the parameters $a_{0}$, $a_{1}$, $a_{2}$ (with $x= a_0 + a_1 z+ a_2 z^2$) of the axial track projection, which are used to trasform $u/v$-coordinates in $y$-coordinates. As for the axial part, the parameters describing the $y-z$ view of the track projection are chosen to be $y_0$ and $y_{11}$, the $y$-coordinates of intersections on the virtual planes of the track associated to the relative pattern cell. Since the fringe magnetic field on the $y-z$ view is very small, the same approximations used for the axial part hold, and are much more accurate. 
Therefore, in order to calculate the retina receptors for the track pattern cells in the $(y_0, y_{11})$ space, the same procedure used to determine receptors for the axial retina is adopted. 

On the contrary to the axial reconstruction, where only a single retina is filled with all hits from axial layers,  different stereo retinas are filled for each axial track candidate. Stereo retinas receive only a small fraction of $u/v$-hits, the only ones compatible with trajectory of the associated axial track candidate. 
For this reason the  granularity of the stereo retinas is much lower than the one used for the axial case, and it is chosen to be equal to $50 \times 50$ pattern cells, where the number of cells mapping the effective diagonal band of interest is 500 (see figure~\ref{fig:retinas}). Each stereo retina covers the whole SciFi acceptance region (top or bottom depending on the chosen quadrant).

Reconstructed track candidates in the stereo view, associated to a given axial track candidate, are found with the same procedure developed for the axial retina, but no requirement is made on the stereo $\chi^2_{\rm S}$ returned from the linearized fit to a straight line. Despite the number of loaded $u/v$-coordinate hits is small, the search of local maxima produces sometimes a number of maxima over threshold which is larger than one. 
The $u/v$-hits combinations with the best $\chi^2_{\rm S}$, for each local maximum, are promoted as possibile  $y-z$ track projections. The one with the best $\chi^2_{\rm S}$ value, over all the local maxima, is promoted as the $y-z$ track projection of the processed axial candidates.   The resulting efficiencies in finding the axial track projection and its stereo projection ($\varepsilon_{\rm AS}$) are shown in table~\ref{tab:efficiencies_3030_solostereo}. They are about 80\% for all track categories. If a minimal requirement on the track momentum is applied efficiencies approach the 90\% level.
The association of $u/v$-coordinate hits is not yet optimal and will be optimized in the future developments. In fact, the probability of misassociating the stereo track counterpart has to be very small, once the most difficult task of the pattern recognition (the axial one) is successfully carried out. Thus, three-dimensional efficiencies and ghost rate are expected to reach a similar level to those obtained for the axial reconstruction.

The tracking boards receiving $u/v$-hits from small stereo angle layers will have to process in parallel an average number of three axial track candidates per event,\footnote{The average number of reconstructed axial track in a single tracking board is indeed 3.} and therefore they
should host approximately three (or more than three, it depends on the capacity of the FPGA chip) identical and independent stereo retinas for a total of ($500 \times 3$) 1500 engines per chip. 
The size of the final system, in order to reconstruct three-dimensional T-tracks using raw hits from the SciFi subdetector, will therefore require approximately $10^5$ pattern cells (processing engines) for solving the pattern recognition using the $x$-coordinate layers, and approximately further $1.2\times 10^5$ pattern cells (processing engines) for the association of the stereo track counterparts. Extrapolating from current available hardware prototypes~\cite{cenci_proc_2017,lazzari_proc_2018} the size of such an envisioned processor seems to be affordable with the already available FPGA chips on the market, for a system to be installed during the LHC LS3 (2025-2026).

\section{Conclusions}
These proceedings present the first performance study of the reconstruction in real time of standalone tracks in the Scintillating Fibre Tracker subdetector achievable with the artificial retina architecture, using fully simulated events at the LHCb~Upgrade (LHC Run~3 and Run~4) conditions. This is a crucial milestone on the path of a future realization of the Downstream Tracker processor for the Future Upgrades of the 
\lhcb experiment.  A tracking performance comparable to that obtainable with the offline reconstruction software is achieved. 

\section*{References}


\end{document}

%% file: lhcb-symbols-def_mjm.tex
\usepackage{ifthen} 
\newboolean{uprightparticles}
\setboolean{uprightparticles}{false} 
\usepackage{xspace} 
\usepackage{upgreek}

\def\lhcb {\mbox{LHCb}\xspace}

\def\MagUp {\mbox{\em Mag\kern -0.05em Up}\xspace}

\ifthenelse{\boolean{uprightparticles}}%
{

 \def\Ppi         {\ensuremath{\uppi}\xspace}

 \def\PDelta      {\ensuremath{\Delta}\xspace}                 
 \def\PXi      {\ensuremath{\Xi}\xspace}                 
 \def\PLambda      {\ensuremath{\Lambda}\xspace}                 
 \def\PSigma      {\ensuremath{\Sigma}\xspace}                 
 \def\POmega      {\ensuremath{\Omega}\xspace}                 
 \def\PUpsilon      {\ensuremath{\Upsilon}\xspace}                 
 
 \def\PB      {\ensuremath{\mathrm{B}}\xspace}                 
                  
 \def\PD      {\ensuremath{\mathrm{D}}\xspace}

 \def\PK      {\ensuremath{\mathrm{K}}\xspace}

 \def\Pb      {\ensuremath{\mathrm{b}}\xspace}                 
 \def\Pc      {\ensuremath{\mathrm{c}}\xspace}

 \def\Pi      {\ensuremath{\mathrm{i}}\xspace}

 \def\Pp      {\ensuremath{\mathrm{p}}\xspace}

 \def\Ps      {\ensuremath{\mathrm{s}}\xspace}

}
{

 \def\Ppi         {\ensuremath{\pi}\xspace}

 \mathchardef\PDelta="7101
 \mathchardef\PXi="7104
 \mathchardef\PLambda="7103
 \mathchardef\PSigma="7106
 \mathchardef\POmega="710A
 \mathchardef\PUpsilon="7107
                  
 \def\PB      {\ensuremath{B}\xspace}                 
                  
 \def\PD      {\ensuremath{D}\xspace}

 \def\PK      {\ensuremath{K}\xspace}

 \def\Pb      {\ensuremath{b}\xspace}                 
 \def\Pc      {\ensuremath{c}\xspace}

 \def\Pi      {\ensuremath{i}\xspace}

 \def\Pp      {\ensuremath{p}\xspace}

 \def\Ps      {\ensuremath{s}\xspace}

}

\makeatletter
\ifcase \@ptsize \relax
  \newcommand{\miniscule}{\@setfontsize\miniscule{4}{5}}
\or
  \newcommand{\miniscule}{\@setfontsize\miniscule{5}{6}}
\or
  \newcommand{\miniscule}{\@setfontsize\miniscule{5}{6}}
\fi
\makeatother

\DeclareRobustCommand{\optbar}[1]{\shortstack{{\miniscule (\rule[.5ex]{1.25em}{.18mm})}
  \\ [-.7ex] $#1$}}

\def\squark    {{\ensuremath{\Ps}}\xspace}

\def\cquark    {{\ensuremath{\Pc}}\xspace}

\def\bquark    {{\ensuremath{\Pb}}\xspace}

\def\pion   {{\ensuremath{\Ppi}}\xspace}

\def\pip    {{\ensuremath{\pion^+}}\xspace}
\def\pim    {{\ensuremath{\pion^-}}\xspace}

\def\kaon    {{\ensuremath{\PK}}\xspace}
  \def\Kbar    {{\kern 0.2em\overline{\kern -0.2em \PK}{}}\xspace}

\def\KorKbar    {\kern 0.18em\optbar{\kern -0.18em K}{}\xspace}

\def\Kp      {{\ensuremath{\kaon^+}}\xspace}
\def\Km      {{\ensuremath{\kaon^-}}\xspace}

\def\KS      {{\ensuremath{\kaon^0_{\mathrm{ \scriptscriptstyle S}}}}\xspace}

  \def\Dbar    {{\kern 0.2em\overline{\kern -0.2em \PD}{}}\xspace}
\def\D       {{\ensuremath{\PD}}\xspace}

\def\DorDbar    {\kern 0.18em\optbar{\kern -0.18em D}{}\xspace}
\def\Dz      {{\ensuremath{\D^0}}\xspace}

\def\Dp      {{\ensuremath{\D^+}}\xspace}

\def\Dstar   {{\ensuremath{\D^*}}\xspace}

\def\Dsp     {{\ensuremath{\D^+_\squark}}\xspace}

\def\Bbar    {{\ensuremath{\kern 0.18em\overline{\kern -0.18em \PB}{}}}\xspace}

\def\BorBbar    {\kern 0.18em\optbar{\kern -0.18em B}{}\xspace}

  \def\Y#1S{\ensuremath{\PUpsilon{(#1S)}}\xspace}

\def\proton      {{\ensuremath{\Pp}}\xspace}

\def\Lz          {{\ensuremath{\PLambda}}\xspace}
\def\Lbar        {{\ensuremath{\kern 0.1em\overline{\kern -0.1em\PLambda}}}\xspace}
\def\LorLbar    {\kern 0.18em\optbar{\kern -0.18em \PLambda}{}\xspace}

\def\to                 {\ensuremath{\rightarrow}\xspace}

\def\AT#1     {\ensuremath{A_{\mathrm{T}}^{#1}}\xspace}           

\def\C#1      {\ensuremath{\mathcal{C}_{#1}}\xspace}                       
\def\Cp#1     {\ensuremath{\mathcal{C}_{#1}^{'}}\xspace}                    
\def\Ceff#1   {\ensuremath{\mathcal{C}_{#1}^{\mathrm{(eff)}}}\xspace}        
\def\Cpeff#1  {\ensuremath{\mathcal{C}_{#1}^{'\mathrm{(eff)}}}\xspace}       
\def\Ope#1    {\ensuremath{\mathcal{O}_{#1}}\xspace}                       
\def\Opep#1   {\ensuremath{\mathcal{O}_{#1}^{'}}\xspace}                    

\newcommand{\tev}{\ifthenelse{\boolean{inbibliography}}{\ensuremath{~T\kern -0.05em eV}}{\ensuremath{\mathrm{\,Te\kern -0.1em V}}}\xspace}
\newcommand{\gev}{\ensuremath{\mathrm{\,Ge\kern -0.1em V}}\xspace}
\newcommand{\mev}{\ensuremath{\mathrm{\,Me\kern -0.1em V}}\xspace}
\newcommand{\kev}{\ensuremath{\mathrm{\,ke\kern -0.1em V}}\xspace}
\newcommand{\ev}{\ensuremath{\mathrm{\,e\kern -0.1em V}}\xspace}
\newcommand{\gevc}{\ensuremath{{\mathrm{\,Ge\kern -0.1em V\!/}c}}\xspace}
\newcommand{\mevc}{\ensuremath{{\mathrm{\,Me\kern -0.1em V\!/}c}}\xspace}
\newcommand{\gevcc}{\ensuremath{{\mathrm{\,Ge\kern -0.1em V\!/}c^2}}\xspace}
\newcommand{\gevgevcccc}{\ensuremath{{\mathrm{\,Ge\kern -0.1em V^2\!/}c^4}}\xspace}
\newcommand{\mevcc}{\ensuremath{{\mathrm{\,Me\kern -0.1em V\!/}c^2}}\xspace}

\def\gsim{{~\raise.15em\hbox{$>$}\kern-.85em
          \lower.35em\hbox{$\sim$}~}\xspace}
\def\lsim{{~\raise.15em\hbox{$<$}\kern-.85em
          \lower.35em\hbox{$\sim$}~}\xspace}

\def\tell1  {TELL1\xspace}
\def\ukl1   {UKL1\xspace}

\def\scifi {SciFi\xspace}



%% file: my-symbols-def.tex
\def\pcie40     {PCIe40\xspace}

\newcommand{\pp}{\proton\proton}

